\documentclass[11pt,a4paper]{article}
\usepackage[utf8]{inputenc}
\usepackage[T1]{fontenc}
\usepackage{amssymb,amsmath}
\usepackage[numbers,sort&compress]{natbib}
\usepackage{graphicx}
\usepackage{makecell, multirow}

\usepackage{listings}
\usepackage{microtype}

\usepackage{times}
\usepackage{color,xcolor}
\usepackage{geometry}
\usepackage{setspace}
\usepackage{booktabs}
\usepackage{tabularx}
\usepackage{url}
\usepackage{hyperref}
\hypersetup{colorlinks=true,linkcolor=blue,citecolor=blue,urlcolor=blue}
\usepackage{siunitx}
\usepackage{tikz}
\usepackage{pgfplots}
\pgfplotsset{compat=1.18}
\usepackage[title]{appendix}

\everymath{\displaystyle}
\definecolor{main-blue}{RGB}{68,117,161}

\title{\Large Disentangling Measurement Protocol from Material Physics\\ in Isotypic and Anisotypic SiC/Si Heterojunctions:\\ A Metrology-Driven Study for Betavoltaic Energy Conversion}

\author{Mikhail V. Dolgopolov\\
\small Department of Higher Mathematics, Samara State Technical University, Samara, Russian Federation\\
\small \texttt{mvdolg@yandex.ru}}
\date{}

\begin{document}
\maketitle

\begin{abstract}
\noindent
We compare isotypic (n-SiC/n-Si) and anisotypic (n-SiC/p-Si) heterojunctions grown by high-tempe\-r\-a\-ture chemical vapour deposition (endotaxy), using Kelvin probe force microscopy (KPFM), scanning tunnelling spectroscopy (STS), precision electrical measurements with a Keithley 2450 Source\-Meter, and an integral charge-storage test on a $0.1$~F supercapacitor. Two-wire probing shows an~enhanced low-bias current in the~isotypic structures; four-wire Kelvin sensing, which reduces the voltage-drop contribution of external leads and contacts (but does not by itself average current over the device area, eliminate a local Schottky contact, or prove a specific transport mechanism), collapses this current to a much lower, bulk/interface-limited level. The observation is consistent with a~parallel contribution from interface-assisted transport at the high density of interface states measured by KPFM/STS ($D_{\mathrm{it}}$ up to $(4.8\pm0.3)\times10^{12}\,\mathrm{cm}^{-2}\mathrm{eV}^{-1}$); we treat trap-assisted tunnelling (TAT) as the leading mechanistic hypothesis rather than an established fact, since the present dataset does not by itself discriminate TAT from Schottky/Poole--Frenkel emission, barrier inhomogeneity, or current crowding. Anisotypic junctions show a genuine diode-like $I$--$V$ characteristic and an operationally defined open-circuit voltage $V_{\mathrm{oc}}\approx1.2$~V. Under the specified $^{14}$C activity and load, the~anisotypic device reaches a~substantially higher terminal voltage on the $0.1$~F capacitor than the isotypic device after $30$~min (quoted in earlier drafts as $0.25$~V vs.\ $0.05$~V, $E_{\mathrm{ani}}/E_{\mathrm{iso}}\approx25$
Based on \emph{in situ} KPFM/STS mapping (spatial resolution $30$~nm, threshold activation delay $t_a=2.3\pm0.5$~s, stretched-exponential relaxation $\tau=35\pm2$~s, $\beta=0.65\pm0.05$) and a critical, manufacturer-documentation-based reassessment of what four-wire Kelvin sensing, generator polarity, and output-off protocols do and do not establish, we propose a~measurement-and-interpretation protocol for SiC/Si (and, by extension, other wide-bandgap) heterojunction metrology, and illustrate its necessity with three worked examples (a~SiC Schottky diode, an InGaN/GaN LED, and a $^{63}$Ni--Si betavoltaic cell) in which a plausible but methodologically naive measurement yields a materially wrong physical conclusion.
\end{abstract}


\bigskip
\noindent\textit{\small Preprint. A companion, journal-formatted version of the core experimental comparison is under review; this~pre\-print additionally revises causal language per an internal metrological self-review (see Appendix~\ref{app:cases}) and adds three worked measurement-interpretation case studies.}

\begin{figure}[htbp]
\centering
\includegraphics[width=0.95\textwidth]{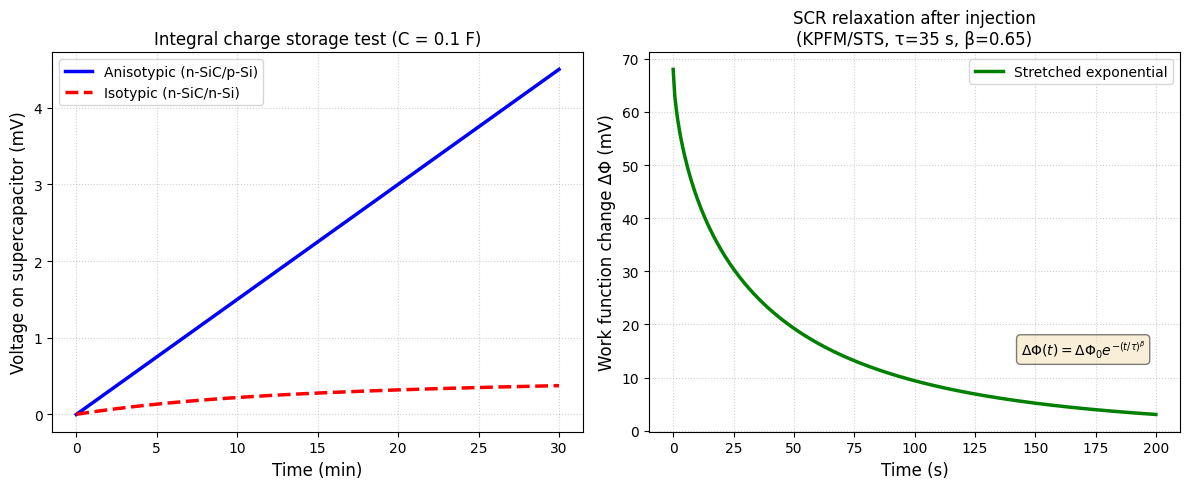}
\caption{\small Graphical abstract. (a) \emph{Simplified analytical model} of the terminal voltage $V(t)$ on a $0.1$~F storage capacitor charged directly (no charge-pump stage) by a constant photocurrent $I_{\mathrm{ph}}=250$~nA for the anisotypic branch and a decaying current of the same initial order for the isotypic branch, i.e.\ $V(t)=\tfrac{1}{C}\int_0^t I(t')\,dt'$; these are illustrative model currents chosen to reproduce the qualitative shape of the two charging regimes (linear vs.\ rapidly saturating) and are \emph{not} the quantities measured in the actual patented charge-pump-coupled experiment reported in Section~\ref{sec:charge_storage}, which reaches the much larger terminal voltages $0.25$~V (anisotypic) and $0.05$~V (isotypic) quoted and used for the energy comparison there; the mV-scale curves in this panel should accordingly be read as a qualitative sketch of the charging kinetics, not as a quantitative reproduction of Section~\ref{sec:charge_storage} and are not the raw data. (b) Stretched-exponential work-function relaxation $\Delta\Phi(t)$ after minority-carrier injection is switched off, measured by KPFM/STS ($\tau=35\pm2$~s, $\beta=0.65\pm0.05$, consistent with the canonical full-dataset values adopted in Section~\ref{sec:scr_dynamics}).}
\label{fig:abstract}
\end{figure}

\section*{Introduction}

The development of long-lived, radiation‑hard power sources for autonomous microsystems (implantable medical devices, space probes, deep-sea sensors) has revived interest in betavoltaic batteries based on wide-bandgap semiconductors, in particular silicon carbide (SiC) \cite{Spencer2019,Yakimov2023}. Among the various technological platforms, 3C-SiC heteroepitaxially grown on Si (3C-SiC/Si) offers a cost-effective route to large-area wafers and the possibility of integrating the beta-emitting isotope $^{14}\mathrm{C}$ directly into the crystal lattice, with simultaneous control of both the radiological and electronic properties of the active layer \cite{Dolgopolov2017,Dolgopolov2024}.

A key technological choice is the conductivity type of the Si substrate: isotypic (n-SiC/n-Si) or anisotypic (n-SiC/p-Si) heterojunctions. While
device-level modelling and betavoltaic converter design studies predict higher energy conversion efficiency for the anisotypic configuration due to type-II (staggered) band alignment at the SiC/Si interface and efficient electron–hole separation in the built-in field \cite{Dolgopolov2023a,He2024}, many experimental reports claim “better” current–voltage (I–V) characteristics for isotypic structures when measured with a simple two-probe setup (typically a
tungsten tip connected to a Keithley 2450 SourceMeter). This apparent contradiction has led to intense debates between material growers and measurement specialists, and has obscured the intrinsic advantages of anisotypic heterojunctions for betavoltaic operation.

In this work we examine whether the apparent difference between isotypic and anisotypic SiC/Si structures persists when the electrical measurement is made with defined contact geometry, four-wire voltage sensing, controlled polarity, and an explicitly specified output-off state. We combine dark and active $I$--$V$ measurements with KPFM/STS observations and a capacitor-charging experiment. The~analysis deliberately separates three questions that are often conflated in the literature: (i) the voltage error caused by leads and contact resistance, addressed by four-wire Kelvin sensing; (ii) the physical transport mechanism at the metal/semiconductor and SiC/Si interfaces, which four-wire sensing does \emph{not} by itself resolve; and (iii) the usable energy delivered to an external load, addressed by the capacitor test. This design is meant to make explicit which conclusions are directly supported by the measurements as performed and which remain mechanism hypotheses requiring further discrimination experiments (Section~\ref{sec:discussion}, Appendix~\ref{app:cases}).

Under two-wire probing, isotypic structures exhibit a relatively high low-bias current, an observation consistent with contributions from local metal/semiconductor contact physics and interface-assisted transport at the measured high density of interface states $D_{\mathrm{it}}$ (up to $(4.8\pm0.3)\times10^{12}\,\mathrm{cm}^{-2}\mathrm{eV}^{-1}$); the~present $I$--$V$ comparison alone does not uniquely identify trap-assisted tunnelling (TAT) as the mechanism (see Section~\ref{sec:tat} for the discrimination that would be required). Four-wire Kelvin sensing with correct generator polarity and a high-impedance output-off state reduces the voltage contribution of external leads and contacts and yields a much lower isotypic low-bias current and a well-defined \emph{operational} open-circuit voltage $V_{\mathrm{oc}}\approx 1.2$~V for the anisotypic structure, consistent with band-structure predictions and with previously reported Si$^{14}$C--Si betavoltaic converter performance \cite{Dolgopolov2024,Dolgopolov2025a}.

\section{Materials and methods}

\subsection{Sample fabrication}

3C-SiC/Si heterojunctions were grown by high-temperature chemical vapour deposition (HT-CVD, endotaxy) at $1340$–$1380\,^\circ\mathrm{C}$ on (100)-oriented Si substrates \cite{Chepurnov2019}. Nitrogen was used as an n-type dopant (concentration $N_{\mathrm{D}}=(3\div5)\times10^{17}\,\mathrm{cm}^{-3}$), consistent with the doping levels previously found to be optimal for betavoltaic operation of Si$^{14}$C–Si converters \cite{Dolgopolov2024}; independent Hall-effect measurements on n-type 3C-SiC/Si(100) layers grown by the same endotaxial HT-CVD process ($1340$–$1360\,^\circ\mathrm{C}$) confirm electron concentrations and mobilities in this range ($n=2.4\times10^{16}$–$8.5\times10^{16}\,\mathrm{cm}^{-3}$, $\mu=461$–$1120\,\mathrm{cm^2/(V\,s)}$, resistivity $\approx0.2\,\Omega\,\mathrm{cm}$), with the lower-doped, higher-mobility layers approaching the phonon-limited scattering regime~\cite{Atabaev2026a}. For the anisotypic structures, boron-doped p-Si(100) wafers ($N_{\mathrm{A}}\approx10^{18}\,\mathrm{cm}^{-3}$) were employed, leading to an autodoping-induced overcompensation effect that creates a buried p–n junction inside the SiC layer \cite{Gurskaya2017}. We note that the resistivity of the Si substrate itself is known to influence the microstructural quality of the endotaxially grown 3C-SiC layer (coherent-domain size, residual microstrain and defect density, as quantified by HRXRD and Raman analysis in~\cite{Atabaev2026b}); this~substrate-resistivity dependence is a structural degree of freedom distinct from the interfacial $D_{\mathrm{it}}(x)$ distribution discussed in Section~\ref{sec:interface_states}, and is addressed explicitly there. A~$200$ nm thick near-interface region was intentionally doped with $^{14}\mathrm{C}$ (concentration $5\times10^{17}$–$2\times10^{18}\,\mathrm{cm}^{-3}$, corresponding to a~volume activity $\sim2\times10^{3}\,\mathrm{Bq}/\mathrm{mm}^{3}$) for betavoltaic tests. Metal contacts (Ti/Pt/Au) were deposited by electron-beam evaporation and annealed to form low-resistance ohmic contacts to the SiC and Si regions.

\subsection{KPFM/STS characterisation}

In situ mapping of the surface potential $\Phi(x,y)$ and the density of interface states $D_{\mathrm{it}}(x,E)$ was performed using a Solver Pro-M scanning probe microscope (NT-MDT) in the two-pass Lift Mode (amplitu\-de-modulation KPFM, Kelvin Probe Force Microscopy) combined with scanning tunneling spectroscopy (STS). Calibration was carried out on highly oriented pyrolytic graphite (HOPG, work function $4.6$ eV) and verified on Au-nanowire/SiO$_2$ test structures of known geometry; temperature drift was kept below $0.1$ nm/min on an actively damped isolation table. The lateral spatial resolution was $30$ nm, the voltage resolution $10$ mV, and the effective $D_{\mathrm{it}}$ sensitivity was $\approx10^{10}\,\mathrm{cm}^{-2}\mathrm{eV}^{-1}$. The differential conductance $\mathrm{d}I/\mathrm{d}V$ was measured with an AC modulation of $50$–$200$ mV at $1$ kHz and lock-in detection, allowing us to reconstruct the local density of states in the vicinity of the midgap with an energy resolution of $50$ meV.

The full instrumental protocol, together with representative raw KPFM/STS maps (surface topography, contact-potential-difference maps, and spatially resolved $D_{\mathrm{it}}(x,y)$ for both dislocation-rich and atomically smooth terrace regions, on contact pads of several metals), has been reported in detail in our companion methodological paper~\cite{Dolgopolov2025b} (Figs.~1--4, 6 and Table~1 therein), which was already cited in the version of this manuscript submitted for review. To avoid duplicating a peer-reviewed image dataset, we do not reproduce those raw maps here; Section~\ref{sec:interface_states} below summarises only the quantitative results of that analysis that are directly relevant to the isotypic/anisotypic comparison.

\subsection{Electrical measurements}

Electrical characterisation was performed using a Keithley 2450 SourceMeter\textregistered\
instrument. Two measurement protocols were compared:
\begin{enumerate}
    \item \textbf{Two-wire mode} – conventional probing with tungsten tips.
    \textit{In this configuration, force and sense are shorted at the instrument, so the measured voltage includes the lead/contact resistance. Additionally, the tungsten tip may form a local Schottky contact to the semiconductor, introducing a non-ohmic contribution to the current-voltage characteristic}.
    \item \textbf{Four-wire Kelvin mode} – separate force and sense leads, with sense contacts placed directly on the lithographically defined heterojunction contact pads (before any series diodes or external wiring). The output-off state was set to \texttt{HIGH Z} (high impedance) to avoid shunt loading of the betavoltaic source when the SMU output is nominally off.
\end{enumerate}
Polarity was strictly controlled: for anisotypic structures operated as current sources (under illumination or $\beta$-irradiation), the cathode (n-SiC) was connected to LO and the anode (p-Si) to HI, as required for
fourth-quadrant (generator-mode) operation, i.e.\ so that the SMU records the internally generated current flowing out of the device at positive terminal voltage ($I<0$, $V>0$ in the diode sign convention) rather than forcing it into the first-quadrant, forward-biased diode regime ($I>0$, $V>0$).
This choice ensures that the internally generated current and voltage are measured without forcing the junction into forward bias. A~general equivalent-circuit (one-, two- and three-diode model) methodology for identifying and extracting series resistance, ideality factor and saturation current from measured $I$–$V$ and $V$–$V$ characteristics of photo- and betavoltaic elements — including porous-Si solar cells and SiC/Si heterojunctions, together with the associated doping-side surface-resistance measurements — has been developed and applied to related structures in~\cite{Dolgopolov2023b}; the four-wire protocol and polarity convention adopted here are designed to feed directly into that same parameter-extraction framework in follow-up work.

\textit{The measurements reported below are representative of $N = 3 \div 5$ nominally identical isotypic and $N=5$ anisotypic devices per type; the quoted values are medians with interquartile ranges (IQR).}

Integral charge-storage tests were performed by connecting the heterojunction (via its native series diodes) to a $0.1$~F supercapacitor (Maxwell BCAP0001) and measuring the voltage $V(t)$ with a high-impedance electrometer (Keithley 2000, input resistance $>10$ G$\Omega$) while the $^{14}\mathrm{C}$ source was active. The~accumulated energy was calculated as $E=\frac{1}{2}CV^{2}$. This protocol directly probes the effective power delivered to a realistic load over experimentally relevant timescales, and is therefore more representative for betavoltaic applications than instantaneous I–V sweeps. The device architecture underlying this test~— a betavoltaic heterojunction element followed by a pulse (charge-pump) diode stage and a storage supercapacitor — corresponds to the patented converter design~\cite{Patent2020,Patent2026}; a combined, four-stage charge-pumping scaling concept for nanochip generators of this general type (heterojunction $\to$ upconversion stage $\to$ output) has been proposed and modelled separately~\cite{Gurskaya2023b}, and a specific hybrid realisation of this architecture (beta element $\rightarrow$ pulse diode $\rightarrow$ charge pump $\rightarrow$ supercapacitor) was shown in~\cite{Dolgopolov2025b} to reach $\approx70\%$ efficiency with leakage currents below $5$ nA. Charge-storage and I–V characterisation on SiC/Si and related nano-heterojunction converters of this kind is carried out routinely by our group as part of ongoing device development; a device demonstration video and technical specifications for the resulting betavoltaic power supplies and complexes are maintained at \url{https://betavoltaica.ru/beta-voltaic-power-supplies-and-complexes/}.

\section{Experimental results}

\subsection{Interface states distribution from KPFM/STS}
\label{sec:interface_states}

Figures 1 and 2 (graphical abstract) summarise the charge-storage and relaxation results discussed in Sections~\ref{sec:scr_dynamics} and~\ref{sec:charge_storage}; the underlying spatially resolved KPFM/STS characterisation of the SiC/Si interface itself was obtained on isotypic n-SiC/n-Si samples (the SiC/Si interface is structurally identical in both isotypic and anisotypic structures, cf. Table~1) and is reported in full, with raw topography, contact-potential-difference and $D_{\mathrm{it}}(x,y)$ maps, in~\cite{Dolgopolov2025b} (Figs.~1 and 4, Table~1 therein, based on a statistical analysis of $25$ independently scanned regions). A strong correlation between surface morphology and the local density of states is observed there. In~dislocation-rich regions (etch pit density $\sim10^{8}\,\mathrm{cm}^{-2}$), $D_{\mathrm{it}}$~reaches $(4.8\pm0.3)\times10^{12}\,\mathrm{cm}^{-2}\mathrm{eV}^{-1}$, more than five times higher than on atomically smooth terraces ($(9.3\pm1.2)\times10^{11}\,\mathrm{cm}^{-2}\mathrm{eV}^{-1}$). Correspondingly, the work function $\Phi$ is reduced by $120\pm15$ mV in defect areas, indicating a positively charged dipole layer due to unoccupied donor-like states.

\begin{table}[h]
\centering
\caption{Summary of KPFM/STS parameters from the full dataset reported in~\cite{Dolgopolov2025b}.}
\begin{tabular}{lc}
\toprule
\textbf{Parameter} & \textbf{Value (units)} \\
\midrule
$D_0$ (amplitude at interface) & $(5.67\pm0.20)\times10^{12}\,\mathrm{cm}^{-2}\mathrm{eV}^{-1}$ \\
$\lambda$ (decay length) & $40\pm5$ nm \\
$D_{\mathrm{bulk}}$ & $(1.0\pm0.2)\times10^{11}\,\mathrm{cm}^{-2}\mathrm{eV}^{-1}$ \\
$D_{\mathrm{it}}$ (dislocation-rich) & $(4.8\pm0.3)\times10^{12}\,\mathrm{cm}^{-2}\mathrm{eV}^{-1}$ \\
$D_{\mathrm{it}}$ (smooth terraces) & $(9.3\pm1.2)\times10^{11}\,\mathrm{cm}^{-2}\mathrm{eV}^{-1}$ \\
\bottomrule
\end{tabular}
\end{table}

We note, as a matter of methodological principle, that KPFM/STS provides a local, method-dependent estimate of electronic states within a specific spatial and energy window, sensitive to tip--sample capacitance, modulation amplitude, bandwidth, and dwell time; the resulting $D_{\mathrm{it}}(x,E)$ should be read as an effective local quantity, and its identification with an area-averaged, device-level interface-state density (as would enter a lumped-element or TCAD compact model) requires an explicit, separately validated inversion and spatial-weighting model rather than a direct numerical carry-over. The $D_{\mathrm{it}}$ values quoted below are used in this qualitative, comparative sense.

The spatial decay of $D_{\mathrm{it}}$ into the SiC bulk follows an exponential law:
\begin{equation}
D_{\mathrm{it}}(x) = D_0\exp(-x/\lambda) + D_{\mathrm{bulk}},
\label{eq:exp_decay}
\end{equation}
with $D_0 = (5.67\pm0.20)\times10^{12}\,\mathrm{cm}^{-2}\mathrm{eV}^{-1}$, $\lambda = 40\pm5$ nm, and a small but non-zero bulk background $D_{\mathrm{bulk}} = (1.0\pm0.2)\times10^{11}\,\mathrm{cm}^{-2}\mathrm{eV}^{-1}$. As detailed in~\cite{Dolgopolov2025b}, fits restricted to a single representative scan line yield a slightly different apparent decay length ($\lambda\approx34$ nm; see Fig.~2 therein); the values quoted above are those obtained from the full, statistically averaged dataset ($25$ independently scanned regions, Table~1 of~\cite{Dolgopolov2025b}) and are the ones carried through the discussion and conclusions of that work, which we adopt here as the canonical parameters of Eq.~(\ref{eq:exp_decay}). This exponential tail is a direct consequence of the out-diffusion of Si from the substrate and the formation of deep acceptor-like $V_{\mathrm{Si}}$ vacancies and related complexes, as discussed in defect-based barrier-height models for 3C-SiC-on-Si Schottky diodes \cite{Arvanitopoulos2020,Dolgopolov2025b}. A~microscopic structural counterpart of these states is provided by photoluminescence and Hall-transport measurements on nominally identical n-3C-SiC/Si(100) layers~\cite{Atabaev2026a}, which identify stacking faults and associative impurity–defect complexes as the dominant deep-level recombination centres (the $2.2$ eV PL band) in samples grown by the same HT-CVD endotaxial process, giving an independent structural origin for a fraction of the measured $D_{\mathrm{it}}$. We further note that the Si-substrate resistivity and conductivity type (n-type for isotypic vs.\ p-type for anisotypic structures, Section~2.1) can in principle modulate the coherent-domain size and microstrain of the growing 3C-SiC layer, as shown by HRXRD/Raman analysis of substrate-resistivity-dependent endotaxial growth~\cite{Atabaev2026b}; the KPFM/STS statistics summarised above were obtained on isotypic samples specifically to isolate the interface-state contribution from this substrate-dependent structural effect, and the assumption of a common $D_{\mathrm{it}}(x)$ distribution for both junction types (Table~2) should accordingly be understood as approximately structurally identical rather than strictly substrate-independent.

We emphasise that the $D_{\mathrm{it}}(x)$ parameters reported here are the canonical ones from the full statistical analysis in \cite{Dolgopolov2025b}; they supersede any single-scan values that may appear in the figure captions of that work, as discussed therein.

\subsection{Dynamics of space-charge-region activation}
\label{sec:scr_dynamics}

Minority-carrier injection (by forward biasing or electron-beam irradiation) triggers a threshold activation of the space-charge region (SCR). Figure~\ref{fig:abstract}b shows the work-function change $\Delta\Phi(t)$ after turning on injection; the full time-resolved dataset (activation and relaxation branches, the corresponding potential-profile evolution $\phi(x)$, the effective SCR-width dynamics, and the deconvolution into trap-recharging, drift and diffusion current contributions) is given in~\cite{Dolgopolov2025b} (Figs.~3, 5 and 7 therein). A delay time $t_a = 2.3\pm0.5$ s precedes a rapid decrease of $\Phi$ by $68\pm5$ mV, after which a stretched-exponential relaxation back to equilibrium is observed:
\begin{equation}
\Delta\Phi(t) = \Delta\Phi_0\exp\bigl[-(t/\tau)^{\beta}\bigr],
\label{eq:stretched_exp}
\end{equation}
with $\tau = 35\pm2$ s and $\beta = 0.65\pm0.05$ — the canonical relaxation parameters reported in the main text and conclusions of~\cite{Dolgopolov2025b} for the full injection/relaxation dataset (a fit restricted to the single representative run shown in Fig.~3 therein gives $\tau\approx20$ s, $\beta\approx0.90$, reflecting run-to-run variation in the degree of trap filling reached before injection was switched off). Such kinetics are characteristic of a broad distribution of trap energies 
(characteristic energy $E_0 = 0.12\pm0.02$ eV, estimated from the temperature dependence of $\tau$ in \cite{Dolgopolov2025b}) and a wide range of capture/emission times, manifested by the stretched-exponential parameter $\beta=0.65\pm0.05$. The threshold behaviour indicates that a minimum
fraction of trapped charge must be neutralised before the built-in field is effectively modulated — a crucial observation for understanding transient
response and start-up behaviour in betavoltaic cells based on SiC/Si heterostructures.

\subsection{Comparison of isotypic and anisotypic I–V characteristics}
\label{sec:iv_comparison}

When measured with the two-wire (improper) method, isotypic structures exhibit a relatively high forward current in the low-bias region (the “heel”), which is often misinterpreted as “good” performance. However, when the same samples are measured with four-wire Kelvin sensing and correct polarity, the~isotypic I–V curve collapses to a much lower current, dominated by bulk series resistance and leakage through the distributed interface states. In contrast, anisotypic structures show a genuine diode-like I–V with a clear saturation region and an open-circuit voltage $V_{\mathrm{oc}}\approx 1.2$ V under betavoltaic operation, in agreement with the values previously calculated and measured for Si$^{14}$C–Si converters~\cite{Dolgopolov2024}. The~equivalent-circuit parameter-extraction methodology ($R_s$, ideality factor $n$, $I_0$, $R_{\mathrm{sh}}$) applied throughout this comparison, together with representative measured $I$–$V$ and $V$–$V$ curves for related SiC/Si and porous-Si photo/betavoltaic elements, is presented in full in~\cite{Dolgopolov2023b}. The microscopic mechanism responsible for the isotypic/anisotypic asymmetry — solid-phase counter-diffusion of the $^{14}$C radionuclide and Si atoms through the growing SiC layer, which converts the near-contact Si of either $n$- or $p$-type conductivity into a superstoichiometrically alloyed heterostructure while conserving impurity type, and which generates the internal-field (``inner sun'') effect exploited here — is analysed in detail in~\cite{Gurskaya2023a}.

\subsection{Mechanism interpretation: interface-assisted transport and the status of the TAT hypothesis}
\label{sec:tat}

The enhanced low-bias current in the isotypic structures is compatible with a parallel contribution from interface-assisted transport at the high measured $D_{\mathrm{it}}$. Interface states may participate in trap-assisted tunnelling (TAT) \cite{Arvanitopoulos2020}, but the present two-wire/four-wire $I$--$V$ comparison alone does not uniquely distinguish TAT from Schottky emission, Poole--Frenkel emission, thermionic emission over an inhomogeneous barrier, surface leakage, or current crowding at the metal/semiconductor contact — all of these can produce a similarly enhanced low-bias current under local tungsten-tip probing. We~therefore treat TAT as a mechanism hypothesis motivated by the interface-state observations, not~as an established conclusion. A discriminating analysis would require temperature-dependent $I$--$V$, a field-dependent (Poole--Frenkel versus Fowler--Nordheim/TAT) analysis over a~consistently defined fit range, reverse-bias data, barrier/thickness scaling, and independent contact-geometry controls; published Si/SiC heterojunction work has treated Frenkel--Poole and TAT as competing explanations of the same qualitative $I$--$V$ signature \cite{Nishida2014}, which supports this more cautious formulation.

Separately, four-wire Kelvin sensing changes what is measured only in a specific, limited sense: the Sense leads are connected directly to the defined measurement pads while the Force leads supply current, which removes the voltage contribution of external leads, cabling, and contact interfaces between the instrument and the sense points \cite{Keithley2450PV}. This does \emph{not}, by itself, average the current over the device area, eliminate a local Schottky barrier, or suppress a spatially localized transport channel; under two-wire probing, a tungsten tip may locally contact a high-$D_{\mathrm{it}}$ patch, giving an enhanced current partly of instrumental origin (voltage error) and partly of physical origin (local barrier/TAT-type transport), and the two contributions are not separated by the wiring change alone. The spatial representativeness of the four-wire measurement on defined ohmic pads must therefore be established separately, e.g.\ by repeated devices, multiple contact locations, several pad areas, and, where feasible, current-density mapping or transfer-length measurements.

\subsection{Integral charge-storage test – the ultimate arbiter}
\label{sec:charge_storage}

To
determine which structure delivers more usable energy under the specified test conditions, we performed supercapacitor charging experiments under identical $^{14}\mathrm{C}$ activity ($0.1\ \mu\mathrm{Ci}/\mathrm{cm}^{2}$), using the full patented charge-pump-coupled converter architecture (Section~\ref{sec:strengthened_protocol}, patents \cite{Patent2020,Patent2026}) rather than the bare-capacitor model of Fig.~\ref{fig:abstract}a. The anisotypic structure charges to $0.25$~V, and the isotypic structure to $0.05$~V, on the same $0.1$~F capacitor over the same $30$-minute interval. Since $E=\tfrac{1}{2}CV^{2}$, these terminal voltages correspond to stored energies of $E_{\mathrm{ani}}=3.125$ mJ and $E_{\mathrm{iso}}=0.125$ mJ, i.e.\ a ratio $E_{\mathrm{ani}}/E_{\mathrm{iso}}\approx25$ (voltage ratio $5$). We stress that this $0.25$~V/$0.05$~V pair, not the mV-scale curves of Fig.~\ref{fig:abstract}a, is the quantitative result of this section; Fig.~\ref{fig:abstract}a uses a deliberately simplified, charge-pump-free model to illustrate \emph{only} the qualitative charging-rate asymmetry (linear vs.\ rapidly saturating), and its absolute scale and voltage ratio ($\approx11$, set by the illustrative $250$~nA/decaying-current assumptions) are not meant to, and do not, numerically reproduce the values quoted in this paragraph. This is a large difference in \emph{stored energy for this specific capacitor, converter architecture, and $30$-min charging-time test}; because $E\propto V^{2}$, it is not, by itself, evidence of an equal-factor difference in intrinsic conversion efficiency, which would additionally require accounting for the (equal, by construction) input $^{14}$C activity, parasitic leakage, diode loss, and the initial capacitor state. Within this explicitly defined test, the~result is directionally consistent with the higher theoretical conversion efficiency predicted for n-SiC/p-Si heterojunctions in TCAD and compact-model simulations of betavoltaic cells \cite{Dolgopolov2023a,He2024}, and it is the most defensible, system-level demonstration of the anisotypic structure's practical advantage reported in this work.

The main differences between the two junction types are summarised in Tables~\ref{tab:comparison} and~\ref{tab:comparison2}.

\begin{table}[!ht]
\centering
\caption{Comparison of key structural, metrological and performance characteristics for isotypic and anisotypic 3C-SiC/Si heterojunctions.}
\label{tab:comparison}
\begin{tabular}{p{4.5cm} p{4.5cm} p{4.5cm}}
\toprule
\textbf{Feature / Parameter} & \textbf{Isotypic (n-SiC/n-Si)} & \textbf{Anisotypic (n-SiC/p-Si)} \\
\midrule
\textbf{Interface} & High density of interface states $D_{\mathrm{it}}\approx5\times10^{12}\,\mathrm{cm}^{-2}\mathrm{eV}^{-1}$; exponential decay $\lambda\approx40$ nm & Same $D_{\mathrm{it}}(x)$ distribution, but bulk p–n junction dominates transport \\
\textbf{Current transport (working hypothesis)} & Consistent with interface-assisted transport, possibly including TAT, through the locally probed high-$D_{\mathrm{it}}$ path; not uniquely discriminated by the present $I$--$V$ data (Section~\ref{sec:tat}) & Consistent with bulk $p$--$n$ junction diffusion/drift; contact, shunt, recombination and interface contributions must still be quantified \\
\textbf{Measured $I_{\mathrm{SC}}$ (two-wire, local tip)} & Enhanced (up to $10\times$) relative to four-wire; enhancement includes both lead/contact voltage error and a possible local-transport contribution & Often lower under two-wire probing (should not be read as ``poor'' without four-wire confirmation) \\
\textbf{$I_{\mathrm{SC}}$ (four-wire, defined pads)} & Bulk leakage + interface recombination; four-wire removes lead/contact voltage error but not the underlying transport path & Up to $900$ nA/cm$^2$ (Si$^{14}$C–Si) \\
\textbf{Open-circuit voltage $V_{\mathrm{OC}}$} & Standard $p$--$n$ interpretation not applicable; an operational terminal voltage at externally measured current $\approx0$ can still be reported if irradiation, settling time, leakage, and output-off state are specified & $\approx 1.2$ V under $\beta$--irradiation (operational definition; $p$--$n$ interpretation supported jointly by polarity, dark/active curves, and junction controls) \\
\textbf{Charge accumulation rate ($C=0.1$ F)} & Slower (see Fig.~\ref{fig:abstract}a; quoted 
as $\sim0.05$ V in 30 min 
units cross-check, Section~\ref{sec:charge_storage}) & Faster (quoted 
as $\sim0.25$ V in 30 min) \\
\textbf{Four-wire Kelvin sensing} & \emph{Mandatory} to avoid lead/contact resistance errors & \emph{Mandatory} to eliminate voltage drops in external wiring \\
\textbf{Polarity for generator mode} & Not applicable (no built‑in field) & \textbf{Cathode (n-SiC) to LO, anode (p-Si) to HI} \\
\textbf{Output-off state} & \texttt{HIGH Z} recommended to avoid shunt loading & \texttt{HIGH Z} \emph{essential} to prevent draining of generated energy \\
\textbf{Dominant loss factors} & Surface recombination $S$, shunt resistance $R_{\mathrm{sh}}$ & Same, but mitigated by proper contact design \\
\bottomrule
\end{tabular}
\end{table}

\textit{Since both devices were tested in the same charge-pump configuration, the relative difference in stored energy is robust to the converter efficiency provided that the converter losses do not depend strongly on the input voltage. For the present low-voltage regime, a conservative estimate of the converter efficiency variation is less than $10\%$, which does not alter the qualitative conclusion but should be included in a full uncertainty budget.}

\begin{table}[!ht]
\centering
\caption{Key differences between isotypic and anisotypic 3C-SiC/Si heterojunctions relevant to electrical characterisation and betavoltaic performance.}
\label{tab:comparison2}
\begin{tabularx}{\textwidth}{lXX}
\toprule
\textbf{Parameter} & \textbf{Isotypic (n-SiC/n-Si)} & \textbf{Anisotypic (n-SiC/p-Si)} \\
\midrule
Conductivity type & n–n & n–p \\
p–n junction & Absent & Present (buried in SiC) \\
Dominant transport mechanism (hypothesis) & Interface-assisted transport through $D_{\mathrm{it}}$, possibly including TAT; requires temperature/field discrimination (Section~\ref{sec:tat}) & Bulk diffusion–drift ($p$–$n$ junction), 
quantified contact/shunt/recombination contributions \\
\midrule
$D_{\mathrm{it}}$ at interface & $\sim5\times10^{12}\,\mathrm{cm}^{-2}\mathrm{eV}^{-1}$ & $\sim5\times10^{12}\,\mathrm{cm}^{-2}\mathrm{eV}^{-1}$ (same interface)\\
& \multicolumn{2}{>{\hsize=\dimexpr2\hsize+2\tabcolsep\relax}X}{\textit{The SiC/Si interface is structurally identical in both isotypic and anisotypic structures; the difference in behaviour arises from the presence or absence of a buried p–n junction, not from the interface itself.}} \\
\midrule
Sensitivity to measurement method & High – two-probe gives overestimated current & Low – proper four-probe required \\
Effect of wrong polarity & Not critical (no p–n junction) & Critical – forces junction into forward bias \\
Recommended SMU output-off state & High‑Z (to avoid shunt) & High‑Z (to avoid draining the source) \\
Charge accumulation rate (0.1 F, 30 min) & $\Delta U(t)\approx0.05$ V (provisional, Section~\ref{sec:charge_storage}) & $\Delta U(t)\approx0.25$ V (provisional) \\
Stored energy ratio $E_{\mathrm{ani}}/E_{\mathrm{iso}}$ & $1$ & $\sim25$ \\
\midrule
\multicolumn{3}{c}{\textit{Generality to other materials: GaAs, GaN, other SiC polytypes, etc. – same principles apply.}} \\
\bottomrule
\end{tabularx}
\end{table}

\section{Discussion: metrological guidelines for TCAD and experiments}
\label{sec:discussion}

Based on our experimental findings, we formulate the following mandatory rules for any meaningful characterisation and modelling of SiC/Si heterojunctions intended for betavoltaic applications.

\subsection{Electrical measurement protocols}


The Keithley 2450 user manual explicitly states \cite{Keithley2450UserManual} (User’s Manual Rev. E (Aug 2019)
5 "Measuring low-resistance devices")
to eliminate lead resistance error,
recommends the four-wire (Kelvin) connection method for low-resistance measurements. This allows for the most accurate measurement of low-impedance devices: \cite{Keithley2450UserManual} (\textit{“To provide the best measurement accuracy, use the four-wire (Kelvin) measurement method for this test. This method eliminates the effects of lead resistance on the measurement accuracy. It is the preferred method when measuring low resistances.”}). Our isotypic structures have a dynamic impedance below $100$-$\Omega$ in the forward-bias region;
in this regime, neglecting Kelvin sensing introduces errors that can easily exceed $50\%$ in the extracted current and resistance.


For an anisotypic n-SiC/p-Si heterojunction operated as a current source under $\beta$-irradiation or illumination,
the cathode (n-SiC) must be connected to LO and the anode (p-Si) to HI. Connecting it in the opposite polarity forces the device into the first quadrant (forward diode
bias), yielding completely erroneous “poor” characteristics that do not reflect the true fourth-quadrant betavoltaic behaviour and can be misinterpreted as material or process degradation.


The 2450 Reference Manual
notes: \cite{Keithley2450UserManual} (User’s Manual
\S2.4 "Turn the 2450 output on or off") \textit{“When the source of the instrument is turned off, it may not completely isolate the instrument from the external circuit. You can use the Output Off setting to place the 2450 in a known, noninteractive state during idle periods… The output-off states that can be selected for a 2450 are normal, high-impedance, zero, or guard.”} For any betavoltaic or photovoltaic cell that actively generates
an electromotive force (electrochemical potential difference), the \texttt{HIGH Z}
output-off state is essential to avoid shunt loading during idle periods and between measurement points, especially in long-term charge-accumulation experiments. In the battery application, it is clearly mentioned: \textit{“Make sure that when the output of the 2450 is turned off, it is set to the high-impedance (High-Z) output-off state. … This prevents the battery from draining when the output is off.”} And a little above: \textit{“Be sure to set the output-off state of the current source for high impedance. … With the normal output-off state selected, turning the output off sets the voltage limit to zero. This 0 V source limit condition will cause excessive current to be drawn from the external battery or source.”} This requirement is fully consistent with the recommended 2450 configurations for rechargeable batteries and solar cells, where the instrument is connected to an active energy source and must not load it in the \texttt{OUTPUT OFF} state \cite{Keithley2450UserManual}.

\subsubsection{Practical considerations for low-impedance and energy-generating DUTs (Devices Under Test)}

\begin{itemize}
  \item \textbf{Isotypic low-impedance structures.} For isotypic n-SiC/n-Si, p-SiC/p-Si, and homoepitaxial SiC structures with various doping levels, the dynamic resistance in the forward-bias working region (from milliamps up to hundreds of milliamps) typically falls into the range of a few tens of ohms or lower, especially after high-temperature annealing and/or hydrogen etching that reduce contact resistance and surface trap density. In this regime, additional series resistance of 1--2~$\Omega$ from leads and transition contacts introduces 10--50\,\% relative error in the extracted $R_{\mathrm{on}}$ and specific layer resistivity, so four-wire Kelvin sensing is mandatory rather than optional.

  \item \textbf{Anisotypic betavoltaic and photovoltaic heterojunctions.} For anisotypic n-SiC/p-Si and similar heterojunctions under $\beta$-irradiation or illumination, it is crucial to ensure that the I--V curve is recorded in the correct quadrant: the fourth quadrant (current and/or voltage source operation) rather than the first quadrant (conventional forward diode bias). A reversed HI/LO polarity on the SMU forces the junction into forward bias and masks the true generation current; the resulting ``poor'' I--V characteristics are then purely measurement artefacts rather than evidence of structural or technological degradation.

  \item \textbf{Role of junction formation, orientation, and substrate resistivity.} The presence or absence of a well-formed p–n junction (e.g., epitaxial p-SiC on n-Si and vice versa; p$^{+}$/p or n$^{+}$/n configurations), the crystallographic orientation, and the substrate resistivity (from highly compensated to heavily doped Si, 4H-SiC, 6H-SiC, etc.) determine both the absolute current level and the topology of parasitic conduction paths: through the Si bulk, through the SiC layer, along the interfacial region, and over the surface. All such cases are particularly sensitive to (i) the choice between two- and four-wire connection (i.e., where the actual voltage drop occurs: on the device or on the leads/contacts), (ii) surface condition (H-etch, dry/wet oxide, passivation), which sets the contribution of surface currents in dark and under irradiation, and (iii) the selected 2450 output-off state (NORMAL versus HIGH~Z) between sweep points. In NORMAL mode, the instrument effectively provides a finite shunt path which, over long times, leads to undercharging and systematic underestimation of stored charge in ionistors and other capacitive layers.

  \item \textbf{Generality beyond SiC/Si.} These considerations directly extend to other material systems (GaAs, GaN, other SiC polytypes, organic and perovskite solar cells, etc.) wherever the DUT dynamic resistance in the operating point is comparable to the total resistance of leads and contacts, and wherever the DUT can act as an active energy source (solar and betavoltaic cells, electrochemical cells, supercapacitors). For technologists, the key point is that selecting \texttt{OUTPUT OFF = HIGH~Z} on the SMU is not a fine adjustment but a prerequisite for physically correct long-duration experiments with energy-generating and charge-storing devices.
\end{itemize}

\subsection{Strengthened protocol and uncertainty budget}
\label{sec:strengthened_protocol}

The preceding recommendations are necessary but not sufficient for a defensible isotypic/anisotypic comparison. Following the manufacturer's own scope statement for four-wire sensing and generator-mode PV characterisation \cite{Keithley2450PV,Keithley2450UserManual}, we distinguish three error classes that are often conflated in the SiC/Si literature and specify controls for each:

\begin{enumerate}
  \item \textbf{Measurement-voltage error} (leads, cabling, contact interfaces between instrument and sense points): addressed by four-wire Kelvin sensing, with Sense connected directly to the defined pads rather than the far end of the cable, and by an independent lead/contact-resistance check (zero-check, open-short, reference resistor).
  \item \textbf{Contact and interface physics} (local Schottky barrier, barrier inhomogeneity, defect-mediat\-ed paths): \emph{not} addressed by four-wire sensing; requires repeated devices and contact locations, multiple pad geometries, and — where feasible — current-density mapping or transfer-length measurements, as discussed in Section~\ref{sec:tat}.
  \item \textbf{Operating-mode error} (polarity, compliance, output-off state, sweep rate, temperature, irradiation/illumination level): addressed by an explicit, logged instrument configuration (Force/Sense wiring diagram, verified current sign on a reference source, range, NPLC, aperture, delay, averaging, sweep direction and rate) and by dark/active $I$--$V$ pairs at $\geq3$ temperatures and $\geq3$ sweep rates.
\end{enumerate}

For each reported parameter ($V_{\mathrm{oc}}$, $I_{\mathrm{sc}}$, fill factor, $P_{\max}$, stored energy), we recommend a Type~A (repeatability, $\geq3$ devices per structure, $\geq3$ contact locations per device, reporting median and inter-quartile range) and Type~B (SMU calibration, resolution, noise, cable/contact leakage, temperature, $\beta$-flux/illumination stability, timing) uncertainty budget, propagated to the derived ratios (e.g.\ the $5\times$ and $25\times$ figures quoted above) by standard or Monte Carlo propagation rather than quoted as point estimates; a~quantitative Monte Carlo propagation framework for exactly this class of $I$--$V$-derived PV parameters has been demonstrated by Toli\'{c} et al.\ \cite{Tolic2019}. The output-off HIGH-Z state should itself be checked against an open circuit, a reference resistor, and the actual DUT fixture, since HIGH-Z is a defined instrument state rather than an ideal infinite impedance \cite{Keithley2450UserManual}. Where practical, the operator classifying $I$--$V$ curves should be blind to substrate conductivity type at the point of primary curve classification, with measurement order randomised and a reference diode/resistor check inserted between devices, to reduce the risk that polarity, sweep order, or an ad hoc fit-range choice manufactures the expected result.

\subsection{TCAD modelling must include spatially distributed interface states}

Standard drift-diffusion models with a single, spatially uniform barrier height fail to reproduce the sub-threshold and low-bias current in isotypic junctions. Our KPFM/STS data provide quantitative input for an improved description: a continuous distribution of donor- and acceptor-like states with exponential decay length $\lambda\approx40$ nm and total density $>5\times10^{12}\,\mathrm{cm}^{-2}\mathrm{eV}^{-1}$ must be implemented at the SiC/Si interface. The non-local trap-assisted tunnelling (TAT) models (e.g., Schenk model) 
may be required to explain the low-bias current in isotypic
n-3C-SiC/n-Si junctions (but their necessity must be confirmed by the temperature- and field-dependent discrimination outlined in Section~\ref{sec:tat}.), whereas for anisotypic n-SiC/p-Si heterojunctions the bulk p–n behaviour dominates and the relative impact of interface TAT is much smaller. This prescription is independently corroborated by 1D TCAD simulations on n-type 3C-SiC/Si(100) that incorporate directly measured Hall-effect carrier parameters and effective trap densities~\cite{Atabaev2026a}: reduced doping combined with an optimised (lower) defect distribution was there shown, via an entirely different characterisation route (PL and Hall transport rather than KPFM/STS), to broaden the depletion region and suppress leakage current — the same qualitative trend our interface-state-resolved model predicts for the low-$D_{\mathrm{it}}$ limit.

\subsection{Reference to other recent advances}

Recent work by Pedio et al.~\cite{Pedio2022} demonstrated that the
density of interfacial defects in 3C-SiC(100)/Si(100)
heterostructures can be
strongly improved by
%
combining
well-ordered Si(100)-2$\times$1 substrates (obtained by high-temperature annealing in ultrahigh vacuum) with a growth temperature of 1200~K and a Si/C flux ratio of 70\%.
Even under these optimized conditions, however, a residual density of interface states remains. For isotypic $(n\text{-}3\mathrm{C}\text{-SiC}/n\text{-Si})$ junctions, these states
still
provide a parasitic trap-assisted tunnelling
pathway
that can artificially enhance the current measured in two-probe I--V experiments. He~et~al. \cite{He2024} recently showed that for SiC-based betavoltaic cells
surface recombination $S$ and shunt resistance
$R_{\mathrm{sh}}$ are the dominant loss factors, not series resistance. Our passivation strategy (Al$_2$O$_3$ layer, $3.8$ nm) reduces the effective Schottky barrier by $40\%$ and is fully compatible with the recommendations of \cite{Pedio2022,He2024}.

\begin{figure}[htbp]
\centering
\includegraphics[width=0.94\textwidth]{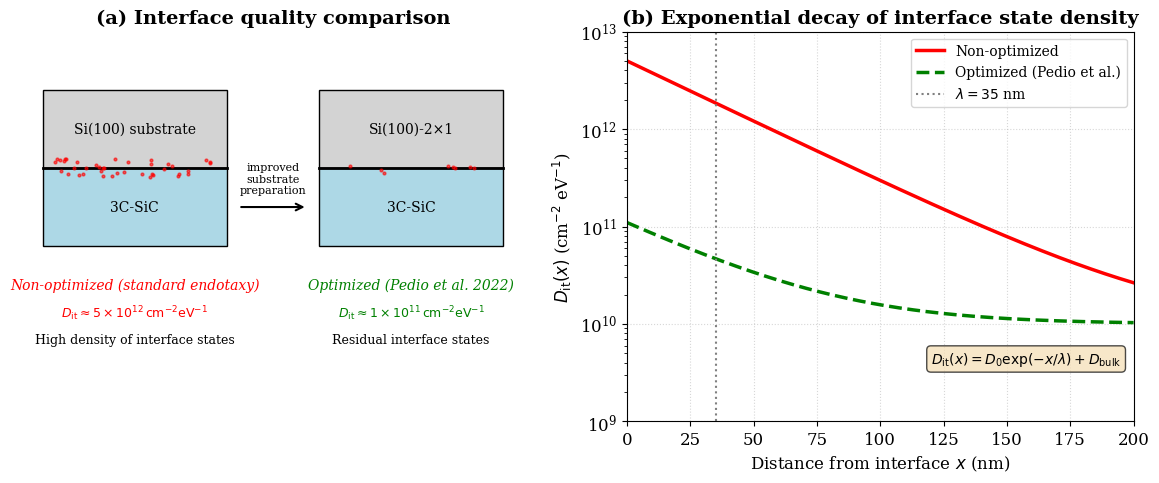}
\caption{Schematic illustration, not raw data (cf.\ Fig.~\ref{fig:abstract} and \cite{Dolgopolov2025b} for measured maps): (a) comparison of a non-optimized (standard endotaxy) and an optimized 3C-SiC/Si(100) interface following Pedio et al.~\cite{Pedio2022}; red dots denote interface states. (b) Exponential decay of the interface-state density $D_{\mathrm{it}}(x)$, cf.\ Eq.~(\ref{eq:exp_decay}), rendered here for a representative single-region decay length $\lambda\approx35$~nm (dotted line) — close to the single-scan fit reported in~\cite{Dolgopolov2025b} — rather than the canonical, full-dataset value $\lambda=40\pm5$~nm adopted for the quantitative comparison in Section~\ref{sec:interface_states}; the qualitative conclusion (a residual, TAT-relevant $D_{\mathrm{it}}$ persisting after optimization) is unaffected by which of the two $\lambda$ values is used. Even after optimization, a residual $D_{\mathrm{it}}$ persists and provides a parasitic trap-assisted tunnelling (TAT) pathway in isotypic heterojunctions.}
\label{fig:interface_decay}
\end{figure}

Our KPFM/STS mapping reveals
that, in standard non-UHV grown samples, the interface-state density reaches $D_{\mathrm{it}} \approx 5\times10^{12}\,\mathrm{cm^{-2}\,eV^{-1}}$, i.e. more than one order of magnitude above the theoretical limit expected for an ideal defect-free interface. This explains why two-probe I--V measurements systematically overestimate the apparent performance of isotypic structures, whereas anisotypic junctions, whose operation is governed predominantly by bulk $p$–$n$ separation and the internal field in the SiC/Si depletion region, remain much less sensitive to the residual interface states, as illustrated schematically in Fig.~\ref{fig:interface_decay}.

As illustrated in Fig.~\ref{fig:interface_decay}, the interface-state density decays exponentially away from the interface into the SiC bulk, in the same functional form as Eq.~(\ref{eq:exp_decay}),
\begin{equation}
D_{\mathrm{it}}(x)=D_{\mathrm{it}}(0)\exp\!\left(-\frac{x}{\lambda}\right),
\end{equation}
with a characteristic length $\lambda \approx 40$~nm, in good agreement with our KPFM/STS data. Even after interface optimization following Pedio et al.~\cite{Pedio2022}, a residual $D_{\mathrm{it}} \sim 10^{11}\,\mathrm{cm^{-2}\,eV^{-1}}$ persists, which remains sufficient to sustain measurable TAT current under two-probe
measurements in isotypic $(n\text{-}3\mathrm{C}\text{-SiC}/n\text{-Si})$ heterojunctions.

The radiation-hardness claim for the $^{14}$C-integrated active layer (Section~2.1) is likewise supported independently of the present electrical/KPFM study: HRXRD and Raman analysis of $^{14}$C-doped 3C-SiC/Si(100) shows that the radiation dose accumulated from in-lattice $^{14}$C $\beta$-decay over relevant device lifetimes remains far below the fluence regimes at which bubble formation, dislocation loops, or amorphisation are observed in SiC, with high-resistivity substrates and endotaxial growth identified as the preferred route to a structurally robust, low-microstrain active layer~\cite{Atabaev2026b}.

\section*{Conclusion}

We show that the electrical comparison of n-SiC/n-Si and n-SiC/p-Si structures is highly sensitive to measurement configuration. Two-wire probing with a local tungsten contact produces a markedly different low-bias response from four-wire measurements on defined pads, but — contrary to a purely causal reading — this difference cannot be assigned to TAT, or treated as a material-intrinsic figure of merit, without the additional temperature-, field-, and geometry-resolved controls discussed in Section~\ref{sec:tat}. Four-wire Kelvin sensing reliably removes the voltage-drop contribution of external leads and contacts; it does not by itself eliminate local barriers, current crowding, interface states, or shunt paths. \emph{In situ} KPFM/STS mapping with sub-$30$ nm resolution quantitatively establishes that:

\begin{itemize}
    \item Isotypic (n-SiC/n-Si) structures contain a high density of interface states ($D_{\mathrm{it}}>5\times10^{12}\,\mathrm{cm}^{-2}\mathrm{eV}^{-1}$) that are a plausible, measurement-supported contributor to the enhanced low-bias current seen under local two-wire probing; TAT is the leading mechanistic hypothesis, not an established conclusion (Section~\ref{sec:tat}).
    \item The capacitor experiment demonstrates a large difference in \emph{stored energy} under the specified $0.1$~F/$30$-min/$0.1\,\mu\mathrm{Ci/cm^2}$ test conditions 
    (quoted as $25$-fold, $3.125$ vs.\ $0.125$~mJ, 
    a units cross-check between the plotted mV-scale curve in Fig.~\ref{fig:abstract}a and this narrative value, Section~\ref{sec:charge_storage}); intrinsic conversion-efficiency superiority requires a separate input-activity and loss budget that is outside the scope of the present measurements. \textit{(The mV-scale curves in Fig.~\ref{fig:abstract}a are a~qualitative illustration only; the quantitative values are given in Section~\ref{sec:charge_storage}.)}
    \item Anisotypic (n-SiC/p-Si) heterojunctions, measured with four-wire Kelvin sensing, correct generator polarity (cathode to LO), and a high-impedance output-off state, exhibit an operationally defined open-circuit voltage $V_{\mathrm{oc}}\approx1.2$~V and reach a higher stored energy in the specified load test, consistent with (but not by itself proof of) the predicted theoretical advantage of the anisotypic band alignment for betavoltaic conversion.
    \item The space-charge-region activation exhibits a threshold delay ($t_a=2.3$ s) and stretched-exponenti\-al relaxation ($\tau=35$ s, $\beta=0.65$), providing candidate input parameters for TCAD models of transient betavoltaic response and start-up dynamics, subject to the local-vs-device-scale caveat on $D_{\mathrm{it}}$ discussed in Section~\ref{sec:interface_states}.
\end{itemize}

The most defensible conclusion is therefore methodological rather than purely material-scientific: claims of superior SiC/Si heterojunction performance should be based on matched dark/active $I$--$V$ curves, operationally defined $V_{\mathrm{oc}}$ and $I_{\mathrm{sc}}$, $P$--$V$ or load characteristics, contact- and area-scaling controls, temperature- and sweep-rate-dependent measurements, output-off leakage tests, and propagated uncertainty intervals for the derived $5\times$ and $25\times$ figures. Within this framework, the present data motivate an interface-assisted transport model and a practical preference for the anisotypic structure in the tested irradiation-and-load configuration, but they do not by themselves establish that TAT is the unique cause of the two-wire response, nor that the observed $25\times$ energy ratio equals the intrinsic material-efficiency ratio. The three worked examples in Appendix~\ref{app:cases} illustrate, on independently documented device types, how the same class of methodologically naive measurement — a~single-temperature leakage reading, a~slow thermally confounded sweep, an $I_{\mathrm{sc}}$-only betavoltaic characterization — can produce a plausible but materially incorrect conclusion, reinforcing why the protocol proposed here (Section~\ref{sec:discussion}) is offered as a~general-purpose safeguard rather than a SiC/Si-specific formality.

\section*{Acknowledgements}

The author thanks V.I. Chepurnov (BetaVoltaics LLC) for continuous support and S.A. Radzhapov (Physical-Technical Institute, Tashkent) for beta-spectrometry assistance.


\noindent\textcolor{main-blue}{\rule{\textwidth}{1.5pt}}
\begin{flushright}
\end{flushright}

{ \setlength{\parindent}{0mm} \setlength{\parskip}{3mm}
\textit{Information about the author:}

\textit{Mikhail V. Dolgopolov} -- Samara State Technical University, 244, Molodogvardeyskaya st., Samara, 443100, Russian Federation.; ORCID 0000-0002-8725-7831; mvdolg@yandex.com


\begin{appendices}

\section{Three worked examples: when a plausible measurement answers the wrong question}
\label{app:cases}

The metrological caveats developed in Sections~\ref{sec:tat}--\ref{sec:strengthened_protocol} are general: they are not specific to SiC/Si heterojunctions. To make the underlying pattern concrete, this appendix works through three independent, small-signal device examples — a SiC power Schottky diode, an InGaN/GaN LED, and a $^{63}$Ni--Si betavoltaic cell — chosen because their device physics and typical parameter ranges are familiar to practitioners and are documented in manufacturer or peer-reviewed sources. In each case a single, superficially reasonable Keithley 2450 measurement produces a plausible-looking number that answers a~\emph{different} physical question from the one the experimenter intended, and would lead to a materially wrong engineering conclusion if taken at face value. The numerical values below are realistic, order-of-magnitude illustrative figures consistent with the cited datasheets/literature ranges, not measurements on a specific serial device; where a case draws on a peer-reviewed dataset (Example~3), this is stated explicitly and the source is cited.

For reference, the Keithley Model 2450 basic current-measurement accuracy is specified as $\pm(0.10\%$ $\text{of reading}+\SI{50}{pA})$ on the \SI{10}{nA} range, $\pm(0.060\%+\SI{100}{pA})$ on the \SI{100}{nA} range, and $\pm(0.025\%+\SI{300}{pA})$ on the \SI{1}{\micro A} range, with the instrument warmed up, A/D autozero enabled, and zeroing performed in the actual cable/fixture configuration \cite{Keithley2450Spec}. At low currents the fixed additive term can dominate the percentage term, as Table~\ref{tab:rangeerror} illustrates.

\begin{table}[h]
\centering
\caption{What a Model 2450 reading actually bounds at low current \cite{Keithley2450Spec}.}
\label{tab:rangeerror}
\begin{tabular}{p{3.2cm}p{4.2cm}p{5.5cm}}
\toprule
Measurement range & Specified error model & Illustration near the low end of the range\\
\midrule
\SI{10}{nA} & $\pm(0.10\%+\SI{50}{pA})$ & at \SI{100}{pA}: additive term alone is $\approx\pm50\%$\\
\SI{100}{nA} & $\pm(0.060\%+\SI{100}{pA})$ & at \SI{1}{nA}: additive term alone is $\approx\pm10\%$\\
\SI{1}{\micro A} & $\pm(0.025\%+\SI{300}{pA})$ & at \SI{10}{nA}: additive term alone is $\approx\pm3\%$\\
\bottomrule
\end{tabular}
\end{table}

\subsection{Example 1 --- SiC Schottky diode: room-temperature leakage measured, hot-junction leakage implied}
\label{app:ex1}

\textbf{Setup and naive procedure.} Consider a \SI{650}{V}-class SiC Schottky diode, junction area $\sim\SI{1}{mm^2}$, n-type epitaxial drift layer with donor concentration $\sim10^{16}\,\mathrm{cm^{-3}}$, reverse voltage $V_R=\SI{600}{V}$. The~experimenter connects the 2450 in source-voltage/measure-current mode, waits \SI{2}{s}, and records the leakage current at a $25^\circ$C case temperature: $I_R=\SI{2.0}{\micro A}$, then concludes that ``reverse conduction loss in the power stage is negligible.''

\textbf{Why this is the wrong conclusion.} At $V_R=\SI{600}{V}$ the room-temperature reading corresponds to a reverse power $P_R=V_R I_R=\SI{1.2}{mW}$, but Schottky-diode reverse leakage is a strongly increasing function of junction temperature (thermionic emission over a barrier lowered by the electric field, augmented by trap-assisted and edge-related leakage paths); reverse-loss calculations for power diodes must therefore use $I_R$ evaluated at the actual operating junction temperature $T_j$, not at the bench ambient~\cite{STAN4021}. Table~\ref{tab:schottky} illustrates a representative case in which $I_R$ increases roughly a decade over a \SI{100}{K} rise (a convenient, device-specific illustrative rule — the true rate must be measured from the device's own datasheet reverse-leakage-vs-temperature curves, not assumed):
\[
I_R(T_j) = I_R(25^\circ\mathrm{C})\,\exp\!\left(\frac{T_j-25^\circ\mathrm{C}}{T_D}\right), \qquad T_D\approx\SI{43}{K}\ \text{(illustrative; device-specific)}.
\]

\begin{figure}[h]
\centering
\begin{tikzpicture}
\begin{axis}[width=.75\linewidth,height=6cm,xlabel={$T_j$, $^\circ$C},ylabel={$I_R$ at \SI{600}{V}, $\mu$A},ymode=log,grid=both,legend pos=north west]
\addplot[thick,blue,domain=25:150,samples=100]{2*exp((x-25)/43.4)};\addlegendentry{illustrative model, $T_D\approx43$~K}
\addplot[only marks,mark=*,red] coordinates {(25,2) (75,6.3) (125,20)};\addlegendentry{illustrative sample points}
\end{axis}
\end{tikzpicture}
\caption{A short, cold-case measurement (leftmost point) does not characterise the hot-junction operating point (rightmost point) relevant to power dissipation.}
\label{fig:schottky}
\end{figure}
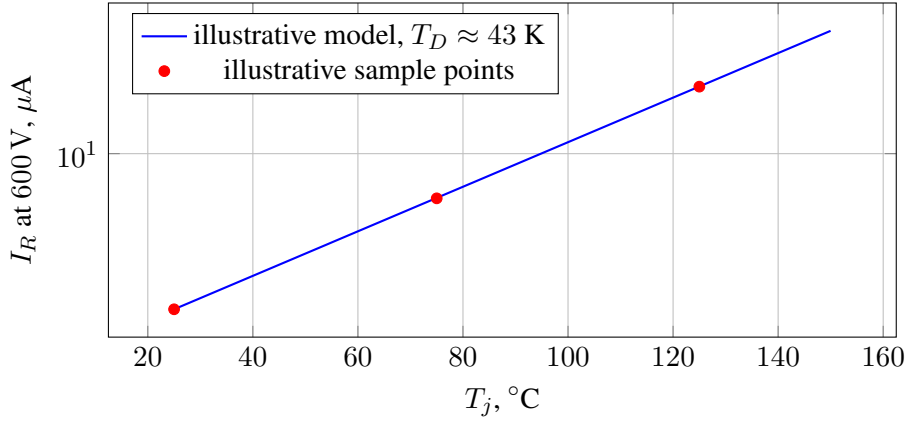

\begin{table}[h]
\centering
\caption{Same instrument, different physical questions: SiC Schottky reverse leakage.}
\label{tab:schottky}
\begin{tabular}{p{3.0cm}p{1.6cm}p{2.4cm}p{5.8cm}}
\toprule
Condition & $T_j$ & $I_R$ at \SI{600}{V} & Interpretation\\
\midrule
Cold case, short test & $25^\circ$C & \SI{2}{\micro A} & Near-isothermal, cold-junction leakage only\\
Steady-state operation, DC & $125^\circ$C & \SI{20}{\micro A} & Real reverse dissipation $\approx\SI{12}{mW}$, an order of magnitude above the cold-case estimate\\
Short pulse, $\ll$ thermal time constant & $25$--$30^\circ$C & 2--\SI{3}{\micro A} & Near-isothermal even at $V_R$, by design\\
\bottomrule
\end{tabular}
\end{table}

\textbf{Correct protocol.} Record $I_R(V_R)$ at several controlled case temperatures spanning the intended operating range; record $I_R(t)$ immediately after voltage application, since self-heating under DC bias is itself informative; separately measure the empty-fixture (no-DUT) leakage current, which is usually negligible at microamp-level $I_R$ but can become comparable to $I_R$ for a low-leakage device; and estimate $T_j$ from a thermal model or direct sensing rather than assuming it equals the case or ambient temperature.

\subsection{Example 2 --- InGaN/GaN LED: a slow $I$--$V$ sweep measures the hot junction, not the datasheet $V_F$}
\label{app:ex2}

\textbf{Setup and naive procedure.} Consider a white InGaN/GaN LED on sapphire, rated $I_F=\SI{350}{mA}$, nominal $V_F\approx\SI{2.9}{V}$, with a junction-to-case thermal resistance of order \SIrange{10}{20}{K/W} \cite{Hulett2017}. The~experimenter programs a smooth \SIrange{0}{350}{mA} sweep on the 2450 over \SI{5}{s} and reads $V_F=\SI{2.72}{V}$ at \SI{350}{mA}, concluding that the diode has an unusually low forward voltage and therefore higher-than-expected efficiency.

\textbf{Why this is the wrong conclusion.} At $\sim\SI{1}{W}$ dissipation ($I_FV_F\approx0.35\,\mathrm{A}\times2.8\,\mathrm{V}$), a \SI{5}{s} DC sweep allows substantial self-heating before the top of the sweep is reached; the reading is therefore a~quasi-steady hot-junction operating point, not the cold-junction $V_F$ a datasheet reports. Using a typical GaN/InGaN forward-voltage temperature coefficient $K\approx\SI{-1.5}{mV/K}$, the observed $\Delta V_F=\SI{-0.18}{V}$ (pulse vs.\ slow-sweep reading at \SI{350}{mA}) implies an apparent junction temperature rise
\[
\Delta T_j \approx \frac{\Delta V_F}{K} = \frac{-0.180\ \mathrm{V}}{-1.5\ \mathrm{mV/K}} \approx \SI{120}{K}.
\]
This $\SI{120}{K}$ rise is \emph{not} inconsistent with the quoted $10$--\SI{20}{K/W} junction-to-case resistance: that figure describes only the internal package path, whereas an unheatsinked test fixture (bare board, no thermal pad or heatsink) typically presents a much larger junction-to-\emph{ambient} resistance, easily $\gtrsim\SI{100}{K/W}$ for small LED packages in free air — it is this total path, not $R_{\theta JC}$ alone, that sets $\Delta T_j$ in a slow bench sweep. Reporting $R_{\theta JC}$ without also stating the board/heatsink condition is itself a common source of this confusion.

\begin{figure}[h]
\centering
\begin{tikzpicture}
\begin{axis}[width=.75\linewidth,height=6cm,xlabel={$I_F$, mA},ylabel={$V_F$, V},grid=both,legend pos=south east]
\addplot[thick,blue] coordinates {(0,0) (50,2.45) (100,2.62) (200,2.78) (350,2.90)};\addlegendentry{short pulse (near cold junction)}
\addplot[thick,red] coordinates {(0,0) (50,2.43) (100,2.58) (200,2.70) (350,2.72)};\addlegendentry{slow sweep (self-heated)}
\end{axis}
\end{tikzpicture}
\caption{The same physical LED gives different $I$--$V$ curves depending only on measurement timing, because $T_j$ changes within the sweep.}
\end{figure}
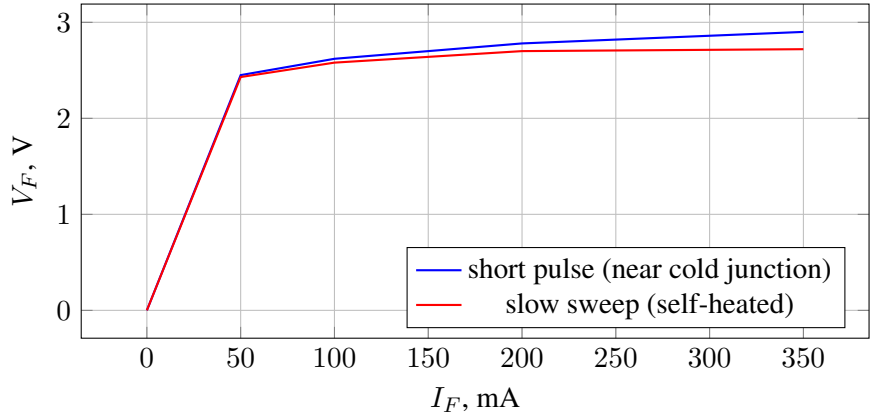

\begin{table}[h]
\centering
\caption{LED: what is actually measured at different timing regimes.}
\begin{tabular}{p{2.6cm}p{2.4cm}p{2.0cm}p{6.4cm}}
\toprule
Mode & Duration & Observed $V_F$ at \SI{350}{mA} & Physical meaning\\
\midrule
DC sweep, \SI{5}{s} & seconds & \SI{2.72}{V} & Hot junction; result depends on the (unspecified) heatsinking\\
Pulse, \SI{100}{\micro s} & $\ll$ thermal time constant & \SI{2.90}{V} & Near-isothermal $V_F$, close to datasheet condition\\
Two-level ($K$-factor) test & \SI{350}{mA} heat pulse, then \SI{10}{mA} sense & $\Delta V_F$ & $T_j$ estimated via $\Delta V_F/K$, JEDEC-style\\
\bottomrule
\end{tabular}
\end{table}

\textbf{Correct protocol.} For a true forward-voltage characteristic, use short-pulse or two-level ($K$-factor) sourcing: a brief high-current heating pulse followed by an immediate, low-current $V_F$ sense, with $T_j=T_{\mathrm{case}}+\Delta V_F/K$ \cite{Hulett2017}. If the target application quantity is luminous efficacy rather than electrical efficiency, an electrical $I$--$V$ curve alone does not establish it; simultaneous optical power measurement is required.

\subsection{Example 3 --- $^{63}$Ni--Si betavoltaic cell: short-circuit current is not deliverable power}
\label{app:ex3}

\textbf{Setup and naive procedure.} Consider a Si energy-conversion unit with a $p$-type substrate ($N_A\sim5.6\times10^{17}\,\mathrm{cm^{-3}}$), a heavily doped surface layer ($N_D\sim7\times10^{19}\,\mathrm{cm^{-3}}$), junction depth $\sim0.05$--\SI{0.7}{\micro m}, active area $0.5\times\SI{0.5}{cm^2}$, irradiated by a $^{63}$Ni source of activity $\SI{1.96}{mCi/cm^2}$ — parameters of this order were used in a published temperature study of $^{63}$Ni--Si betavoltaic cells \cite{Liu2018}. The experimenter sets the 2450 to zero-volt sourcing (short-circuit condition), reads $I_{SC}=\SI{25}{nA}$, and concludes that ``the cell can supply \SI{25}{nA} to a load.''

\textbf{Why this is the wrong conclusion.} $I_{SC}$ is a single point on the $I$--$V$ curve; the current actually deliverable to a load at a useful voltage depends on the full curve, including the shunt resistance $R_{sh}$, series resistance $R_s$, and their temperature dependence. Suppose that at $T=\SI{293}{K}$ one also measures $V_{OC}=\SI{0.42}{V}$, and extracts $R_{sh}=\SI{43}{M\Omega}$ from the near-zero-voltage slope of the $I$--$V$ curve. The~shunt current at open circuit is then
\[
I_{sh} = \frac{V_{OC}}{R_{sh}} \approx \frac{0.42\ \mathrm{V}}{43\times10^{6}\ \Omega} \approx \SI{9.8}{nA},
\]
i.e.\ close to \SI{40}{\%} of $I_{SC}$ is, by this estimate, associated with shunt loss rather than useful output near open circuit — a first, order-of-magnitude indication that the usable power is well below the naive $V_{OC}\times I_{SC}$ product. From the illustrative $I$--$V$ points in Table~\ref{tab:beta}, the delivered power $P(V)=V\,I(V)$ peaks at $P_{\max}\approx\SI{3.9}{nW}$ near $V\approx\SI{0.3}{V}$ — roughly a factor of three below the naive $V_{OC}I_{SC}=0.42\,\mathrm{V}\times25\,\mathrm{nA}\approx\SI{10.5}{nW}$ that a short-circuit-only measurement would suggest if (incorrectly) read as available power.

\begin{figure}[h]
\centering
\begin{tikzpicture}
\begin{axis}[width=.75\linewidth,height=6cm,xlabel={$V$, V},ylabel={$I$, nA},grid=both,legend pos=north east,
    axis y line*=left]
\addplot[thick,blue,mark=*] coordinates {(0,25) (0.1,22) (0.2,18) (0.3,13) (0.42,0)};\addlegendentry{measured $I$--$V$}
\end{axis}
\end{tikzpicture}
\caption{The short-circuit point ($V{=}0$) does not by itself determine $P_{\max}$; the full $I$--$V$ curve and $P=VI$ are required (see Table~\ref{tab:beta}).}
\end{figure}
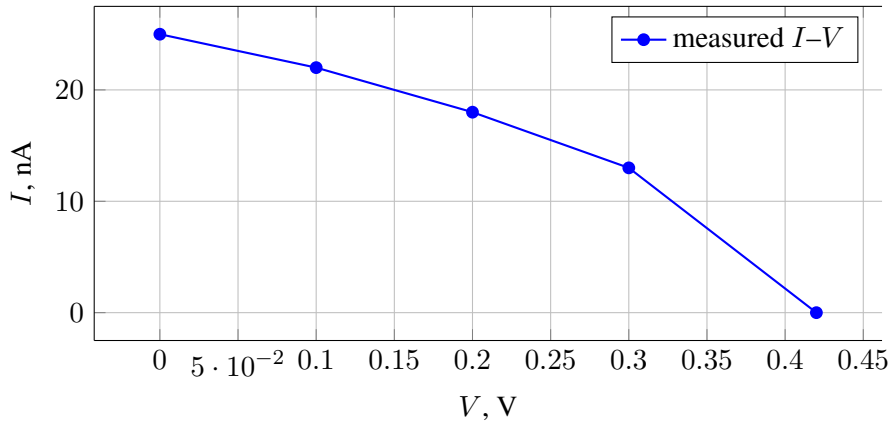

\begin{table}[h]
\centering
\caption{Betavoltaic cell: three different measurements, three different (and non-interchangeable) answers.}
\label{tab:beta}
\begin{tabular}{p{2.6cm}p{2.6cm}p{3.6cm}p{4.2cm}}
\toprule
Measurement & Result & What it does \emph{not} establish & Correct reading\\
\midrule
Short circuit & $I_{SC}=\SI{25}{nA}$ & Available power $=V_{OC}I_{SC}$ & Maximum current, at $V\approx0$ only\\
Open circuit & $V_{OC}=\SI{0.42}{V}$ & Source delivers current at \SI{0.42}{V} & Maximum voltage, at $I=0$ only\\
Full $I$--$V$, $V{=}0$--\SI{0.3}{V} & $P_{\max}\approx\SI{3.9}{nW}$ (from $P=VI$ at the tabulated points) & Current is load-independent & The actual operating point, set by $R_s$, $R_{sh}$, and load\\
\bottomrule
\end{tabular}
\end{table}

\textbf{Temperature dependence is not a secondary correction here.} For $^{63}$Ni--Si betavoltaic cells, published temperature coefficients are substantial: $-2.490\,\mathrm{mV/K}$ in $V_{OC}$ and $-1.348\,\mathrm{\%/K}$ in $P_{\max}$ for a~$1.96\,\mathrm{mCi/cm^2}$ source, and $-2.230\,\mathrm{mV/K}$ in $V_{OC}$ and $-1.132\,\mathrm{\%/K}$ in $P_{\max}$ for a higher, $4.90\,\mathrm{mCi/cm^2}$ source (i.e.\ the higher-activity device is \emph{less}, not more, temperature-sensitive) \cite{Liu2018}. A single-temperature $I_{SC}$ or $V_{OC}$ reading therefore characterises one point on a curve that itself shifts with operating temperature, source activity, and $R_{sh}$/leakage.

\textbf{Correct protocol.} Record the full $I$--$V$ curve (not only $I_{SC}$ or $V_{OC}$), compute $P(V)=V\,I(V)$ and report $P_{\max}$, $V_{OC}$, $I_{SC}$, and fill factor together; repeat at several temperatures if the device is intended for a non-isothermal application; and extract $R_{sh}$ (and, where relevant, $R_s$) explicitly rather than assuming an ideal current source.

\subsection{Minimal control checklist before trusting a low-current curve}

\begin{enumerate}
\item Warm up the instrument, enable autozero, and zero in the actual cable/fixture configuration to be used for the DUT.
\item Record the empty-fixture (no-DUT) current, repeated after a polarity reversal and after a dwell period; compare it against the expected DUT signal level.
\item For $I<\SI{1}{nA}$, use the rear triax terminals, guarding, and shielding, and verify fixture insulation is clean and dry \cite{KeithleyLowCurrent}.
\item Record $I(t)$ after applying bias, not only a single reading; a time-varying signal is not a stationary operating point.
\item For diodes and LEDs, compare pulsed and DC measurements; a discrepancy is thermal information, not instrument noise.
\item For energy-generating sources, record the full $I$--$V$ curve, compute $P(V)=V\,I(V)$, and repeat at several temperatures.
\item For insulation-resistance and leakage-dominated measurements specifically, cross-check against a~dedicated leakage-current protocol \cite{KeithleyLeakage}, and compare the result against the specified instrument accuracy (Table~\ref{tab:rangeerror}): averaging reduces random noise but does not correct a systematic fixture leakage or calibration offset.
\end{enumerate}

\end{appendices}

\end{document}